\documentclass[%
 preprint,
showkeys,
nofootinbib,
nobibnotes,
 amsmath,amssymb,
 aps,
pra,
 longbibliography,
 lengthcheck,%
]{revtex4-1}

\usepackage{graphicx}
\usepackage{dcolumn}
\usepackage{bm}
\usepackage{hyperref}

\newcommand{\be}{\begin{equation}}
\newcommand{\ee}{\end{equation}}

\newcommand{\la}{\langle}
\newcommand{\ra}{\rangle}

\newcommand{\cH}{{\cal H}}
\newcommand{\cS}{{\cal S}}
\newcommand{\cB}{{\cal B}}

\newcommand{\hH}{ H}

\newcommand{\hHS}{ H^{\cal S}}

\newcommand{\densS}{ \rho^{\cal S}}

\newcommand{\meandensS}{\overline{{ \rho}^{\cal S}}}

\newcommand{\trB}{\mathrm{tr}_{\cal B}}

\newcommand{\im}{\mathrm{i}}

\usepackage{amsthm}

\theoremstyle{definition}
\newtheorem*{question}{Question}

\theoremstyle{definition}
\newtheorem*{answer}{Answer}

\theoremstyle{definition}
\newtheorem*{answer cite}{Answer \cite{zeh1973toward}\cite{zurek1982environment}}

\theoremstyle{theorem}
\newtheorem*{lemma}{Lemma}

\theoremstyle{definition}
\newtheorem*{EDH}{Eigenstate Decoherence Hypothesis (EDH)}

\begin{document}

\preprint{APS/123-QED}

\title{Dependence of decoherence-assisted classicality on the ways a system is partitioned into subsystems}

\author{Oleg Lychkovskiy}
\affiliation{%
 Lancaster University, Physics Department, UK,
}%
\affiliation{%
Institute for Theoretical and Experimental Physics, Moscow, Russia.
}%


\begin{abstract}
Choosing a specific way of dividing a closed system into parts is a starting point for the decoherence program and for the quantum thermalization program. It is shown that one can always chose such way of partitioning that decoherence-assisted classicality does not emerge and thermalization does not occur. Implications of this result are discussed.


\end{abstract}

\maketitle


\section{Introduction}

Consider a large closed quantum system with a state space $\cH$ and a Hamiltonian $H.$ It is ubiquitous in physics to resolve
this large closed system in two parts, system $\cS$ and bath $\cB,$\footnote{Throughout the paper we refer to $\cS$ as simply a "system" without a prefix "sub-", while $\cH$ is referred to as a "closed system". Bath $\cB$ is often referred to as "environment".} and to concentrate on the dynamics of $\cS,$ i.e. on the evolution of the reduced density matrix $\densS(t).$  Assume both $\cS$ and $\cB$ are macroscopic. Then, based on the everyday experience, one expects certain features of the reduced evolution to manifest at certain timescales. At a very short time scale $\tau$ the state of the system $\cS$ is expected to {\it decohere}. This means that $\densS(t)$ can be diagonalized in a basis with each basis vector being "quasiclassical" \cite{zeh1973toward}, i.e. providing vanishing uncertainty to every observable from a certain "classical" set (e.g. center-of-mass coordinate, momentum, energy etc). To derive such type of short-time behavior from realistic Hamiltonians is the goal of the {\it decoherence program}, which was essentially founded by  Zeh \cite{zeh1970interpretation,zeh1973toward} and Zurek \cite{zurek1981pointer,zurek1982environment} (see e.g. a book \cite{schlosshauer2008decoherence} for an extensive overview and profound list of references). Decoherence, when proved, accounts for one aspect of quantum-to-classical transition, namely, the apparent absence of Schr\"{o}dinger cat states in our everyday experience (and does it irrespectively to the chosen {\it interpretation} of quantum theory).

At a very large time scale $T$ one expects that the system thermalizes, which means, roughly speaking, that $\densS(t)$ approaches an equilibrium density matrix which depends only on some coarse-grained properties of the bath state, but not on the initial state of the system.  We refer to the effort to justify this intuitive statement from the first principles of quantum theory as a {\it quantum thermalization program}. In the past decade a considerable progress was made in this direction (see e.g. 
\cite{tasaki1998quantum,popescu2006entanglement,goldstein2006canonical,reimann2008foundation,linden2009quantum}; a profound list of references can be found in \cite{gogolin2010pure}).

From a formal point of view,  resolution into subsystems (partitioning) is described by a {\it tensor product structure} (TPS) over the state space $\cH$:
\be
\cH=\cS\otimes\cB.
\ee
Zanardi pointed out that any state space $\cH$ with a nonprime dimension supports infinitely many TPS and that this simple mathematical fact (which is referred to as {\it partitioning relativity}) has profound consequences to quantum physics \cite{zanardi2001virtual}. In particular, he argued that a notion of entanglement is relativized: for any state which is entangled with respect to some TPS one can find another TPS in which this state is factorized.

The goal of the present paper is to show that an analogous conclusion is valid both for decoherence and for thermalization: for every closed quantum system with a state space $\cH$ and a Hamiltonian $H$ there exists a TPS in which  decoherence and thermalization are absent. This will be proven by explicitly constructing such a TPS. It should be emphasized that it is decoherence, not thermalization,  which is of principal interest for us. However we start from thermalization (which is a simpler issue) in order to introduce several important notions to be used when discussing the decoherence. Our results on thermalization are closely related to findings of ref. \cite{gogolin2011absence}.

The ambiguity in resolving a closed system into subsystems was previously realized to be an important conceptual problem for the decoherence program \cite{zurek1998decoherence,omnes2002decoherence}. The present result sharpens this conclusion to the extreme.

Recently Dugi\'{c} and Jekni\'{c}-Dugi\'{c} also considered the implications of the partitioning relativity to the decoherence program \cite{dugic2012parallel}. They studied the system of coupled harmonic oscillators (quantum Brownian motion model) and came to a striking conclusion that within different TPS the decoherence leads to the emergence of qualitatively different classical worlds. We do not discuss this outstanding claim in the present paper. We only stress that it is based on the analysis of a specific model, while our arguments are completely general.

The plan of the rest of the paper is as follows. In Sec. \ref{sec partitionong relativity} we discuss the partitioning relativity, i.e. the freedom to choose different TPS.  We start from an example which motivates consideration of different TPS. 
Then we show how an arbitrary basis in $\cH$ generates a set of TPS.
After that we construct a TPS generated by the eigenbasis of $H.$

Sec. \ref{sec thermalization} is devoted to thermalization. First we discuss what properties of the reduced density matrix constitute thermalization. Then we show that within the eigenbasis-induced TPS one of such properties --- initial state independence of the equilibrium density matrix --- is absent. In the last subsection we discuss the Eigenstate Thermalization Hypothesis (ETH) \cite{deutsch1991quantum,srednicki1994chaos,rigol2008thermalization} and demonstrate that it can not be valid within the considered TPS.

In Sec. \ref{sec decoherence} we turn to the decoherence program. First we specify the notion of the decoherence-assisted  emergent classicality within the context of the present paper. After that we formulate a condition which ensures that the classicality fails to emerge. Then we demonstrate how to construct a TPS within which this condition is fulfilled. This new TPS shares some common features with the eigenbasis-induced TPS, in particular, it also precludes thermalization. In Sec. \ref{subsec EDH} we introduce an Eigenstate Decoherence Hypothesis in a close analogy with the Eigenstate Thermalization Hypothesis and discuss its robustness against partitioning relativity.

In Sec. \ref{sec discussion} the implications of the results are discussed.
In Sec. \ref{sec summary} we summarize our the findings.

\section{\label{sec partitionong relativity}Partitioning relativity}

\subsection{\label{subsec example}Example: partitioning relativity in $XX$ spin chain}

Let us start from a simple but enlightening example. Consider a chain of spins $1/2$ with the $XX$  Hamiltonian:
\be
\hH =\frac14  \sum_{n=1}^{N-1} (\sigma_n^x \sigma_{n+1}^x+  \sigma_n^y \sigma_{n+1}^y)  + \frac{h}2  \sum_{n=1}^N \sigma_n^z
\ee
After Jordan-Wigner and Bogolyubov transformation applied to operators $\sigma_n^{\pm}$ this model appears to be a free-fermion one \cite{lieb1961two},
\be
\hH=\sum_p \varepsilon_p (c_{p}^+ c_{p}-\frac12).
\ee
An operator $c_p^+$ creates a fermion mode with quasi-momentum $p$ (admitting $N$ discrete values) and energy $\varepsilon_p.$ 

What are the possible ways to partition the closed system under consideration into a two-level system  and a $2^{N-1}$-level bath? Evidently, one can regard any spin in the chain as a system $\cS$ and other $N-1$ spins as a bath $\cB$. This gives $N$ possible choices of partitioning. However these $N$ choices by no means exhaust all possibilities. Another set of possible bipartitions emerges when one regards a given fermion mode as a system and other fermion modes as a bath. In fact, there are infinitely many such sets of bipartitions generated by various transformations applied to a set of operators $\sigma_n^{\pm}$. This is what we call {\it partitioning relativity}.

Different ways of partitioning lead to drastically different behavior of corresponding reduced states. If a single spin is regarded as $\cS,$ its reduced density matrix relaxes to an equilibrium one up to finite-size effects, at least for a certain class of initial conditions \cite{abraham1970thermalization} (in case when $\dim S$ is small decoherence and thermalization timescales ar usually of the same order, $\tau\sim T,$ therefore one should not separate a relaxation process into decoherence and thermalization \cite{lychkovskiy2011entanglement}). On the contrary, if a single fermion mode is regarded as $\cS,$ its reduced density matrix does not evolve with time at all, since $c_{p}^+ c_{p}$ is an integral of motion. Properties of entanglement in the $XX$ model also drastically depend on how the closed system is resolved into parts \cite{fel2012quantum}.

\subsection{\label{subsec constructing TPS}Constructing a tensor product structure}

For simplicity, we consider a finite-dimensional Hilbert space $\cH.$ Its dimension $d$ is considered to be a nonprime number:
$d=mn,$ $n>m\gg 1.$ Central for our discussion is the following

\begin{lemma}

Consider an arbitrary basis in $\cH$ with $\dim \cH =mn,~m\geq2,~n\geq 2.$ Enumerate it by a double index $(i,j):$
\be\nonumber
\{\Psi_{i,j}\}, ~~~i=1,2,...,m;~~j=1,2,...,n.
\ee
This enumeration generates a bipartition $\cH=\cS\otimes\cB$ of the space $\cH$ in a tensor product of two spaces $\cS$ and $\cB,$ $\dim\cS=m,~\dim\cB=n,$ such that every basis state $\Psi_{i,j}$ is a product state with respect to this bipartition,
\be\label{identification}
\Psi_{i,j}=\varphi_i \otimes \chi_j,~~\varphi_i\in\cS,~~\chi_j\in\cB,
\ee
and, moreover, $\{\varphi_i\}$ and $\{\chi_j\}$ are bases in $\cS$ and $\cB$ correspondingly.
\end{lemma}
\noindent
Note that all the bases discussed throughout the paper are orthonormal.

The lemma essentially states that every basis in the state space $\cH$ of a closed quantum system generates a TPS (in fact, combinatorially many different TPS).
The statement of the lemma is almost self-evident: one formally introduces Hilbert spaces   $\cS$ and $\cB$ with orthonormal bases $\{\varphi_i\}$ and $\{\chi_j\}$ correspondingly, takes a tensor product $\cS\otimes\cB$, maps the basis $\{\varphi_i\otimes\chi_j\}$ to the basis $\{\Psi_{i,j}\}$ of $\cH$ and thus introduces the identification (\ref{identification}).

\subsection{\label{subsec eigenbasis-induced TPS}Eigenbasis-induced tensor product structure}

Let us apply the above lemma to the eigenbasis $\{|E_l\ra\}, ~ l=1,2,...,d$ of the Hamiltonian $H$.\footnote{We always use Dirac bra-ket notations for states of the closed system $|E_l\ra,$ $|E_{i,j}\ra$ and $|\tilde E_l\ra$ (see below) to avoid the confusion between states and eigenenergies. On the other hand, we often omit bras and kets for the states $\varphi$ and $\chi$ of the subsystems to simplify notations.} To do this one has to introduce a double index $(i,j)$ instead of a single index $l,$ which amounts to arranging basis states in a table:
\begin{table}[h]
\caption{\label{TPS-1}}
\begin{ruledtabular}
\begin{tabular}{c|cllcccccccl}
 &  & $|\chi_1\ra$ & $|\chi_2\ra$ &  &  & .  & . & . &  &  & $|\chi_n\ra$ \\
\hline
 &  &  &  &  &  &  &  &  &  &  &  \\
$|\varphi_1\ra$ &  & $|E_1\ra$ & $|E_2\ra$ &  &  & .  & . & . &  &  & $|E_n\ra$ \\
 &  &  &  &  &  &  &  &  &  &  &  \\
$|\varphi_2\ra$ &  & $|E_{n+1}\ra$ & $|E_{n+2}\ra$ &  &  & .  & . & . &  &  & $|E_{2n}\ra$ \\
 &  &   &  &  &  &  &  &  &  &  &  \\
... &  &  ... & ... &  &  & . & . & . &  &  & ...  \\
 &  &   &  &  &  &  &  &  &  &  &  \\
$|\varphi_m\ra$ &  & $|E_{(m-1)n+1}\ra$ & $|E_{(m-1)n+2}\ra$ &  &  & .  & . & . &  &  & $|E_{mn}\ra$ \\
\end{tabular}
\end{ruledtabular}
\end{table}
\newline
This table defines a map from the set of eigenstates $\{|E_l\ra\}$ to the set of product states $\{\varphi_i \otimes \chi_j\}.$ Combining these basis vectors in superpositions
allows to extend this map to a map between the original state space $\cH$ and the product $\cS\otimes\cB$ of state spaces of two subsystems $\cS$ and $\cB$,  $\{\varphi_i\}$ being a basis in $\cS$ and $\{\chi_j\}$ --- in $\cB.$  For example, the following identifications are in order:
\be
\begin{array}{l}
|E_2\ra  = \varphi_1 \otimes \chi_2, \\
|E_2\ra+|E_n\ra=\varphi_1 \otimes (\chi_2+\chi_n), \\
|E_2\ra+|E_{2n}\ra=\varphi_1 \otimes \chi_2+\varphi_2 \otimes\chi_n.
\end{array}
\ee
In what follows we often enumerate the vectors of the eigenbasis by a double index explicitly:
\be
|E_{i,j}\ra\equiv |E_{(i-1)n+j}\ra=\varphi_i\otimes\chi_j.
\ee
We refer to the TPS defined by Table \ref{TPS-1} as TPS-1.

Note that the eigenvalues  $E_l$ do not have to be arranged in an ascending order. Different orderings of the sequence of eigenvelues  induce different TPS.

A very special feature of TPS-1 which will show up in the next section is that (1) every Hamiltonian eigenstate $|E_l\ra$ is of the product form with respect to TPS-1 and, moreover,  (2) to construct such product states one needs only vectors from a single basis in $\cS$ and a single basis in $\cB.$

In general, TPS-1 corresponds to a very unnatural bipartition. However in certain toy models TPS-1 appears to be natural, for example, in the paradigmatic central spin model with a Hamiltonian
\be\label{central spin model}
H=\sigma_{\rm central}^z \sum_{n=1}^N g_n \sigma_n^z,
\ee
where a central spin $1/2$ is regarded as a system and  $N$ peripheral spins $1/2$ as a bath. This model was studied by Zurek  \cite{zurek1982environment} who demonstrated that the decoherence singles out states $|\uparrow\ra,$ $|\downarrow\ra$ to be pointer states and leads to the fast decay of Schr\"{o}dinger-cat-like  superpositions $(\alpha |\uparrow\ra+\beta |\downarrow\ra)$ to incoherent mixtures. Thus, eigenbasis-induced TPS is not in general suitable for demonstrating the {\it failure} of the decoherence-assisted emergence of classicality. However in the next section we show that the thermalization is absent in TPS-1, while in Sec. \ref{sec decoherence} we turn to a more sophisticated TPS to deal with the decoherence.

\section{\label{sec thermalization}Thermalization}

\subsection{\label{subsec ingridients of thermalization}Ingredients of thermalization}

Define a time-averaged reduced density matrix:
\be
\meandensS \equiv \lim\limits_{t \to \infty}t^{-1}\int_{0}^t \densS(t')dt'.
\ee
It plays a central role in the quantum thermalization program as it encapsulates the long-time behavior of $\cS.$ In particular, if $\densS(t)$ equilibrates, the equilibrium density matrix is $\meandensS$ \cite{linden2009quantum}. As will be clear from Sec. \ref{sec decoherence}, $\meandensS$ is also of substantial importance for the decoherence program.
Throughout the paper we suppose for simplicity that the spectrum of $H$ is nondegenerate.
In this case
\be
\meandensS = \sum_{l=1}^{d} \left|\la E_l| \Psi(0)\ra\right|^2 \trB | E_l\ra\la E_l|,
\ee
where $| \Psi(0)\ra\in\cH$ is an initial state.

What kind of long-time behavior of $\densS(t)$ do we expect from our everyday experience with subsystems?  We expect that $\densS$ approaches an equilibrium density matrix of some special (e.g. Boltzmann-Gibbs) form. As  was argued in \cite{linden2009quantum}, on closer examination one expects that the system exhibits four distinct properties, which we refer to as {\it thermalization properties:}
\begin{enumerate}
    \item{{\it Equilibration.} System $\cS$ is said to {\it equilibrate} if $\densS(t)$ approaches a time-averaged density matrix $\meandensS$ and stays close to it most of the time. Defined in this way, equilibration does not imply neither any special form of $\meandensS,$ nor the independence of $\meandensS$ from initial conditions.}
    \item{{\it Bath initial state independence (Bath ISI).} This means that $\meandensS$ does not depend on the exact initial microstate of the bath. Rather $\meandensS$  depends on some macroscopic characteristics of the state of the bath.
        The prime example of such characteristic is the energy of the bath (which is related to the bath temperature when bath itself is in (quasi-)equilibrium).}
    \item{{\it System initial state independence (System ISI).} This means that $\meandensS$ does not depend on the initial state of the system $\cS$.}
    \item{{\it Boltzmann-Gibbs form of the equilibrium state:} $\meandensS=Z^{-1}\exp(-\beta \tilde H^{\cS}),$ where $\tilde H^{\cS}$ is some effective Hamiltonian of the system $\cS$. This property may be expected if the interaction between the system and the bath is in some sense "weak" and the initial state of the bath has a small energy uncertainty.}
\end{enumerate}
The last three properties make sense only if the firsts one holds. The last property makes sense if also the properties (2) and (3) hold. Note the lack of the symmetry between the definitions of the bath ISI and the system ISI. This asymmetry arises because the bath is assumed to be much larger than the system, so that it can absorb or inject any amount of energy from or to the system and to completely wipe out any memory about the initial state of the system.

A number of theorems were proven in \cite{linden2009quantum} which allow to formulate the above listed thermalization properties in mathematically rigorous terms and to justify the equilibration and the Bath ISI properties under certain conditions imposed on the spectrum of the total Hamiltonian (but not on the TPS). We are in a position to demonstrate that within an eigenbasis-induced TPS-1 the system ISI property does not hold.

\subsection{\label{subsec braekdown of thermalization}Breakdown of the thermalization}

Throughout the paper we consider initial states of a product form (with respect to a chosen TPS):
\be
\Psi(0)=\varphi(0)\otimes\chi(0),~~\varphi(0)\in\cS,~~\chi(0)\in\cB.
\ee
Let us first focus on a very special initial state. Assume that the initial state of the bath is described by a basis vector, $\chi(0)=\chi_j$ (while the initial state $\varphi(0)$ of the system is still arbitrary).
In this case $\densS(t)$ evolves as it were a state of an isolated system with an effective Hamiltonian
$\hHS_j\equiv\sum\limits_i E_{i,j} |\varphi_i\ra\la \varphi_i |:$
\be
\densS(t)=e^{-\im \hH^\cS_j t}|\varphi(0)\ra\la \varphi(0)| e^{\im \hH^\cS_j t}.
\ee
Evidently, the evolution is unitary and does not display any of the thermalization properties (it also does not display the decoherence).

It can be argued, however, that the above observation is not of practical significance since $\chi_j$ is a very special, almost "improbable" initial state of the environment. In practice we do not "prepare" environment (since its microstate is largely uncontrollable), but rather pick by random a state which is restricted only by some global constraints (e.g. the total energy should be in a certain narrow window). This idea is advocated and developed in \cite{linden2009quantum}.    Arguably, a more reasonable setup is to consider a {\it generic} initial state of the bath. Following \cite{linden2009quantum}, we say that some property is valid for a generic initial state of the bath if it is valid for most states from some large linear subspace $\cB_{ R}\subset\cB$ (with possible exceptions forming a set of an exponentially small in $\dim \cB_{R}$ measure).

It was shown in \cite{linden2009quantum} that  under certain reasonable conditions equilibration and  bath ISI properties are valid for an arbitrary initial state  of the system and a generic initial state of the bath.

However, the equilibrium state strongly depends on the initial state $\varphi(0)$ of the system, which is a consequence of the Hamiltonian eigenstate factorizability  with respect to the TPS-1. This follows from a general theorem which relates the breakdown of the system initial state independence with the degree of factorizability of the energy eigenstates \cite{gogolin2011absence}. In our case it is straightforward to demonstrate the lack of the system ISI directly. Indeed,

\be\label{meandensS TPS-1}
\meandensS=\sum_i |\la\varphi_i|\varphi(0)\ra|^2 |\varphi_i\ra\la\varphi_i|.
\ee
Evidently, one can obtain {\it any} set of probabilities $|\la\varphi_i|\varphi(0)\ra|^2$ in the equilibrium reduced density matrix $\meandensS$ choosing suitable initial state.

One can understand the breakdown of thermalization from another perspective. Assume the initial state of the system is $\varphi(0)=\varphi_i$ while the the initial state of the environment $\chi(0)$ is arbitrary (a case opposite to one discussed in the beginning of the present subsection). Then the reduced state of the system does not evolve at all, $\densS(t)=\densS(0)=|\varphi_i\ra\la\varphi_i|,$ and thermalization is absent. Note, however, that the decoherence-assisted classicality still can emerge, as is exemplified by the central spin model (\ref{central spin model}) \cite{zurek1982environment}. In fact, it does emerge for generic spectrum of the total Hamiltonian, in which case states $\varphi_i,~i=1,2,...,m$ form the pointer state basis, while superpositions of these states decay into incoherent mixtures due to interaction with the environment. Thus TPS-1 precludes thermalization but can support the decoherence-assisted classicality.

\subsection{\label{subsec  ETH}Eigenstate Thermalization Hypothesis}

The {\it Eigenstate Thermalization Hypothesis} (ETH)  \cite{deutsch1991quantum,srednicki1994chaos,rigol2008thermalization} is worth mentioning in the present discussion, as it underlies an important branch of the quantum thermalization program. It states that for realistic Hamiltonians thermalization occurs at the eigenstate level:
\be
\trB |E_l\ra\la E_l|\simeq \densS_{\rm eq},
\ee
where $\densS_{\rm eq}$ depends not on the eigenstate itself, but on the value of one or few functionals on $\cH.$ In the simplest and most  physically relevant case this functional is just the total energy: $\densS_{\rm eq}=\densS_{\rm eq}(\la E_l|H|E_l\ra).$
ETH is a strong assumption.
If the energy uncertainty of the initial state is small, ETH guarantees Bath ISI and System ISI. If in addition the interaction between the system and the bath is weak, ETH also guarantees the Boltzmann-Gibbs form of the equilibrium state. One can easily see, however, that the ETH trivially fails within TPS-1. Indeed,
\be
\trB |E_{i,j}\ra\la E_{i,j}|=|\varphi_i\ra\la\varphi_i|,
\ee
which strongly depends on the exact microstate.




\section{\label{sec decoherence}Decoherence}

\subsection{\label{subsec classicality}Decoherence-assisted classicality}
A wave function of a closed system $\Psi(t)$ can be represented in a Schmidt form,
\be
\Psi(t)=\sum_{i=1}^n \sqrt{p_i(t)}~ \varphi_i(t)\otimes\chi_i(t)
\ee
where $\{\varphi_i(t)\}$ form a basis in $\cS$ and $\{\chi_i(t)\}$ can be completed to a basis in $\cB.$
This corresponds to the diagonal form of the reduced density matrix
\be
\densS(t)=\sum_{i=1}^n p_i(t) |\varphi_i(t)\ra\la\varphi_i(t)|.
\ee
States $\varphi_i(t)$ describe alternatives which an observer inside the system $\cS$ is able to perceive (this is true in any interpretation of quantum theory). However the majority of states in  $\cS$  are Schr\"{o}dinger-cat-like states which can not be interpreted classically. What forces $\varphi_i(t)$ to lie in a small quasiclassical domain? The key insight of the decoherence theory is that even if initially $\varphi_i(t)$ are Schr\"{o}dinger-cat-like states, after a very short decoherence time $\tau$ they become quasiclassical states, as a result of interaction between the system and the environment (if the Hamiltonian is nontrivial enough) \cite{zeh1973toward}. We will refer to this dynamical process as the {\it decoherence-assisted emergence of classicality}.

The following subtlety should be emphasized. Assume that for a given  closed system, total Hamiltonian and bipartition the decoherence-assisted classicality does emerge. Consider two reduced density matrices, $\densS$ and $\tilde\rho^{\cS},$ corresponding to different moments of time (such that the decoherence have already occurred) and/or to different initial conditions. They can be diagonalized in the bases $\{\varphi_i\}$ and $\{\tilde \varphi_i\}.$ By assumption, all states  $\varphi_i$ and $\tilde \varphi_i$ belong to a quasiclassical domain of the state space. However, in general it is {\it not} true that bases
$\{\varphi_i\}$ and $\{\tilde \varphi_i\}$ coincide or even are close to each other. The reason is that for sufficiently large dimension of $\cS$ any basis represents a very fine-grained set of alternatives, while classical world is coarse-grained \cite{zeh1970interpretation}. It can appear, for example, that $\tilde\varphi_1=\frac1{\sqrt2}(\varphi_1+\varphi_2).$ Indeed, think of  $\varphi_1$ being a state of an alive cat, with a spin of a single nucleus in its body pointing up, and $\varphi_2$ being a state of alive cat with a spin of that nucleus pointing down. Then $\tilde\varphi_1$ is a state of the alive cat with a spin of the nucleus pointing in $x$ direction -- still a classical state. Thus we do not expect that $\{\varphi_i\}$ and $\{\tilde \varphi_i\}$ coincide, but rather that each state from these two bases belongs to a quasiclassical domain of $\cS$ (cf. \cite{omnes2002decoherence}). How to define this domain?

This question is not easy to answer in general. A usual route is to require that the quasiclassical states have vanishing quantum uncertainties for a set of observables  used in classical mechanics (center-of-mass coordinate, momentum, energy etc). This is not completely satisfying as a priori classical notions are used (a clear discussion of this issue is given in  Sec. VIII B of paper \cite{omnes2002decoherence}). Moreover, this is not directly applicable in our setting, as we allow arbitrary partitioning in which these a priori classical observables in general are not related to the system $\cS$  (in other words, corresponding self-adjoint operators are not of the product form $A^{\cS}\otimes\openone^{\cB}$ ).  One should employ a similar strategy but avoid using a priori classical notions. This strategy can be sketched as follows.

One defines a set $\{A_\alpha\}$ of (dimensionless) observables in $\cS$  which are considered to be classical {\it by definition}. Then one postulates that the states $\varphi\in\cS$ which provide small uncertainties for all $A_\alpha$ (i.e. $\forall \alpha$ $\la\varphi|A^2_\alpha|\varphi\ra-\la\varphi|A_\alpha|\varphi\ra^2<\varepsilon$ with some small $\varepsilon$) form a quasiclassical subset $\cS_{\rm cl}\subset\cS.$ If the decoherence-induced classicality emerges, in the above-discussed sense, with respect to the defined subset $\cS_{\rm cl}$ of  quasiclassical states, then the whole construction is {\it self-consistent} \cite{omnes2002decoherence}. Note that the set $\{A_\alpha\}$  should be sufficiently large and nontrivial to allow for a nontrivial classical reality (this excludes e.g. the set consisting of one element, identity operator) but sufficiently coarse-grained to allow the decoherence to proceed (this excludes e.g. the complete set of $m$ mutually orthogonal projection operators).

\subsection{\label{subsec failure of classicality}Sufficient condition for classicality failure}

We do not follow the above-described route in the present paper. As soon as our goal is modest --- to demonstrate the {\it failure } of the decoherence-induced classicality within {\it some} TPS, we instead identify a condition which signifies such a failure.

Consider two bases in $\cS,$ $\{\varphi_i\}$ and $\{\tilde \varphi_i\},$ related to each other by a discrete Fourier transform (call them {\it maximally distant bases}):
\be\label{Fourier}
|\tilde \varphi_i\ra=\frac1{\sqrt{m}} \sum_{k=1}^{m} e^{-2\pi\im (k-1)i/m} |\varphi_k\ra.
\ee
We argue that the vectors of these two bases can not simultaneously belong to $\cS_{\rm cl}.$ Indeed, assume that each $\varphi_i$ does belong to $\cS_{\rm cl}.$ Although some of the pairs $(\varphi_i,\varphi_{i'})$ can describe the states which are hardly distinguishable classically and therefore can form a superposition which is still classical (such as the states of alive cat with two different directions of a single nuclear spin), a bulk of such pairs should necessarily represent classically distinguishable states (such as states of alive and dead cat), otherwise the classical reality would be trivial. Thus each state $\tilde \varphi_i$ is a superposition of classically distinguishable states, therefore it is highly nonclassical.
\footnote{In the continuous case (which we do not discuss in detail) the eigenbases of momentum and position are maximally distant. Every vector in each of two bases is manifestly nonclassical as momentum eigenstates have infinite position uncertainty and vice versa.}

Let us now consider the equilibrium density matrix $\meandensS.$ An evident necessary requirement for the decoherence-assisted classicality to emerge is that $\meandensS$ describes a mixture of quasiclassical states for any  initial conditions. If, on the contrary, there exist such a partitioning in which  $\meandensS$ can be diagonal in each of the maximally distant bases, depending on the initial conditions, the decoherence-assisted classicality fails to emerge. This is a sought-for sufficient condition for the classicality failure.

\subsection{\label{subsec TPS-2}TPS within which classicality fails to emerge}

Now we are in a position to construct such a TPS in which the above formulated condition is fulfilled.

Assume that $n$ is even. Consider the following basis in $\cH:$
\be
|\Psi_{i,j}\ra=
\left\{
\begin{array}{lll}
|E_{i,j}\ra & {\rm for} & j \leq n/2\\
|\tilde E_{i,j}\ra & {\rm for} & j \geq n/2+1\\
\end{array}
\right.
,
\ee
where $|\tilde E_{i,j}\ra$ is an superposition of $m$ eigenstates:
\be
|\tilde E_{i,j}\ra=\frac1{\sqrt{m}} \sum_{k=1}^{m} e^{2\pi\im (k-1)i/m} |E_{k,j}\ra.
\ee
This basis along with a chosen way of ascribing double indices defines a TPS (call it TPS-2):
\newline
%
\begin{table}[h]
\caption{\label{TPS-2}}
\begin{ruledtabular}
\begin{tabular}{c|cllcccccccl}
 &  & $|\chi_1\ra$ & $|\chi_2\ra$ &...  & $|\chi_{n/2}\ra$ & $|\chi_{n/2+1}\ra$  & ... &  &  &  & $|\chi_n\ra$ \\
\hline
 &  &  &  &  &  &  &  &  &  &  &  \\
$|\varphi_1\ra$ &  & $|E_{1,1}\ra$ & $|E_{1,2}\ra$ & ... & $|E_{1,n/2}\ra$ & $|\tilde E_{1,n/2+1}\ra$ & ... &  &  &  & $|\tilde  E_{1,n}\ra$ \\
 &  &  &  &  &  &  &  &  &  &  &  \\
$|\varphi_2\ra$ &  & $|E_{2,1}\ra$ & $|E_{2,2}\ra$ & ... & $|E_{2,n/2}\ra$ & $|\tilde E_{2,n/2+1}\ra$  & ... &  &  &  & $|\tilde E_{2,n}\ra$ \\
 &  &   &  &  &  &  &  &  &  &  &  \\
... &  &  ... & ... &  & ... & ... & ... &  &  &  & ...  \\
 &  &   &  &  &  &  &  &  &  &  &  \\
$|\varphi_m\ra$ &  & $|E_{m,1}\ra$ & $|E_{m,2}\ra$ & ... & $|E_{m,n/2}\ra$ & $|\tilde E_{m,n/2+1}\ra$  & ... & &  &  & $|\tilde E_{m,n}\ra$ \\
\end{tabular}
\end{ruledtabular}
\end{table}
%
\newline
The first $n/2$ columns of Table \ref{TPS-2} coincide with those of the Table \ref{TPS-1} which defines  TPS-1. The state space of the environment can be represented as a direct sum, $\cB=\cB_1\oplus\cB_2,$ $\cB_1$ and $\cB_2$ being linear hulls of $\{\chi_1,...,\chi_{n/2}\}$  and $\{\chi_{n/2+1},...,\chi_n\}$ correspondingly.  The equilibrium density matrix reads
\begin{eqnarray}\label{meandensS TPS-2}
\meandensS=\left(\sum_{j=1}^{n/2} \left|\la \chi_j |\chi(0)\ra\right|^2\right) \sum_i \left|\la \varphi_i |\varphi(0)\ra\right|^2 |\varphi_i\ra\la\varphi_i|\nonumber\\
+\left(\sum_{j=n/2+1}^{n} \left|\la \chi_j |\chi(0)\ra\right|^2\right) \sum_i \left|\la \tilde\varphi_i |\varphi(0)\ra\right|^2 |\tilde\varphi_i\ra\la\tilde\varphi_i|,
\end{eqnarray}
where the bases $\{\varphi_i\}$ and $\{ \tilde\varphi_i\}$ are related by a Fourier transform (\ref{Fourier}) and thus are maximally distant. If the initial state of the environment $\chi(0)$ belongs to $\cB_1,$ the equilibrium density matrix is diagonal in the basis $\{\varphi_i\}$, while if $\chi(0)\in\cB_2,$ then $\meandensS$ is diagonal in $\{ \tilde\varphi_i \}$. This means that the decoherence-assisted classicality can not emerge.

The above demonstrated property of $\meandensS$ is nothing else than the lack of the bath initial state independence of decoherence (compare to the corresponding property of thermalization). Two subspaces $\cB_1$ and $\cB_2$ of the state space of the environment have large dimensions, namely $n/2$ each. Therefore we have verified the breakdown of the bath ISI property of decoherence for  {\it generic} initial states of the bath.

It is clear from eq. (\ref{meandensS TPS-2}) that if an initial state of the environment has nonvanishing components both in  $\cB_1$ and $\cB_2,$ then the diagonal basis of $\meandensS$ depends also on the initial state $\varphi(0)$ of the system. However the fact that the coefficients $\la \varphi_i |\varphi(0)\ra$ and $\la \tilde\varphi_{i'} |\varphi(0)\ra$ are not independent complicates the situation. We do not consider this issue in detail.

If one considers initial states of the bath from the subspace $\cB_1$  only, then TPS-2 coincides with TPS-1. Therefore the breakdown of thermalization established in the previous section for TPS-1 also occurs for TPS-2.

Let us clarify the difference between TPS-1 and TPS-2.
While all eigenvectors are factorizable with respect  both to TPS-1 and to TPS-2, in the latter case the corresponding products can not be constructed using vectors from a {\it single} basis in $\cS.$\footnote{Tensor product structures with the same property naturally emerge in more sophisticated versions of a central spin model \cite{cucchietti2005decoherence,lychkovskiy2010necessary}.} Indeed, within TPS-2 one needs two maximally distant bases $\{\varphi_i\}$ and $\{\tilde\varphi_i\}$:
\be\label{eigenstates in TPS-2}
|E_{i,j}\ra=
\left\{
\begin{array}{lll}
\varphi_i\otimes \chi_j & {\rm for} & j \leq n/2\\
\tilde\varphi_i\otimes \chi_j & {\rm for} & j \geq n/2+1\\
\end{array}
\right.
\ee
Thus none of the bases  $\{\varphi_i\}$, $\{\tilde\varphi_i\}$ can be a pointer basis. The robustness of a given state of a {\it system} against decoherence depends on the state of the {\it environment}. This situation can be informally depicted as follows. Assume one runs a Schr\"odinger experiment with a cat in a box, the latter being immersed in a heat bath. When the experiment is repeated at different temperatures of the bath, one discovers that at some temperatures the outcomes are either $|\mathrm{alive}\ra$ or $|\mathrm{dead}\ra$ cat, while at other temperatures the outcomes are either $(|\mathrm{alive}\ra+|\mathrm{dead}\ra)/\sqrt{2}$ or $(|\mathrm{alive}\ra-|\mathrm{dead}\ra)/\sqrt{2}.$ This is certainly {\it beyond} the realm of classicality.

\subsection{\label{subsec EDH}Eigenstate Decoherence Hypothesis}

In analogy with the eigenstate thermalization hypothesis discussed in Sec \ref{subsec  ETH}, one can put forward an
\begin{EDH}
 For realistic Hamiltonians  the decoherence-assisted classicality emerges at eigenstate level, i.e. all states $\trB |E_{l}\ra\la E_{l}|$ belong to $\cS_{\rm cl}$.
\end{EDH}

EDH, if valid, would guaranty that the equilibrium density matrix is diagonal in a quasiclassical basis regardless of the initial states of the system and of the environment. However it again appears that this hypothesis can  not be valid within an arbitrary bipartition. Indeed, it is violated within the TPS-2, as is clear from eq. (\ref{eigenstates in TPS-2}).

We believe, however, that the eigenstate decoherence hypothesis can be valid when the the choice of TPS is restricted by some physically motivated requirement (e.g. involving locality). This will be discussed elsewhere.

It is worth stressing that  EDH is a weaker assumption then ETH. For example, it is easy to verify that in a  central spin model (\ref{central spin model}) the ETH fails, in accordance with the general result of Sec. \ref{subsec  ETH}, but EDH holds.


\section{\label{sec discussion}Discussion}


One of the important difficulties beyond the quantum-to-classical transition can be outlined by the following
\begin{question}
Due to superposition principle all states of Hilbert space are equally real. Why in everyday life we observe only tiny fraction of them, but never observe Schr\"{o}dinger-cat-like states, which constitute the vast majority of Hilbert space? How are the quasiclassical states chosen from a Hilbert space?
\end{question}
\noindent
The decoherence program gives the following
\begin{answer}
The only thing the superposition principle guarantees is that every state  of Hilbert space can be (in principle) {\it prepared}. However as long as we deal with an open system interacting with an environment, the state of the former, $\densS(t),$ evolves in such a way that Schr\"{o}dinger-cat-like states decay almost instantly,
and we are left with the quasiclassical states robust against decoherence \cite{schlosshauer2008decoherence}.
\end{answer}
\noindent
However our results motivate another
\begin{question}
The above answer is true for some tensor product structures (in other words, for some ways of resolving a closed system into subsystems) but {\it necessarily  not true} for other tensor product structures. All tensor product structures are equally real. How is a ``right'' TPS chosen from the space of all TPS?
\end{question}
\noindent
This question is in general unanswered \cite{zurek1998decoherence,omnes2002decoherence} (and very often even unaddressed) in the decoherence program.

A pragmatic approach to resolving the partitioning ambiguity was advocated in \cite{zanardi2001virtual,zanardi2004quantum}: the choice of a specific TPS was argued to be "induced by the experimentally accessible observables" \cite{zanardi2004quantum}. This approach, while being reasonable and natural in the context of quantum information processing and other applications, is unsatisfactory from the fundamental point of view \cite{omnes2002decoherence}. Indeed, it is the declared {\it goal} of the decoherence program to decide whether any given  observable is experimentally accessible or not, the decision being based solely on the underlying Hamiltonian.

Anyhow, it seems that the only conceivable way out
is to admit that not all TPS are "equally real": some {\it fundamental} physical requirement distinguishing a certain class of acceptable tensor product structures should be explicitly specified in the decoherence program (as well as in the quantum thermalization program).

Of course, it is clear from causality arguments based on the finiteness of the speed of communication  that there exists a class of distinguished "good" partitionings which respect locality in some form. We emphasize, however, that our arguments are completely independent from this line of reasoning: they are equally valid in a nonrelativistic setting.

Let us make a final brief remark on locality requirement.
At first sight, one could take a straightforward approach: to equip quantum theory with a notion of spatial locality (for example as is described in the book \cite{haag1996local}) and to admit only {\it local} (i.e. confined in bounded regions of three-dimensional space) subsystems. This prescription, advocated e.g. in \cite{wallace2010quantum}, can face several difficulties. We mention here one of them.
The prescription does not cover a common situation in which both thermalization and decoherence are well established \cite{joos1985emergence}: a particle or a macroscopic object regarded as the system interacts through collisions with a gas of host particles regarded as the environment. In this case the particle or the object can explore unbounded regions of space. This example illustrates that a straightforward application of the locality requirement can be too restrictive and thus not completely satisfying. It remains an open question how the partitioning ambiguity  should be resolved in the decoherence and quantum thermalization programs.



\section{\label{sec summary}Summary}

Let us highlight the major results of the present work. Our principle goal was to prove the following statement: for any large closed quantum system a partitioning always exists with respect to which the decoherence-assisted classicality fails to emerge.
To achieve this goal it was necessary, first, to identify a sufficient condition for such a failure which would not rely on any {\it a-priory} classical concepts (in particular, would not involve notions of {\it a-priory} classical, natural or experimentally accessible observables), and, second, to construct a tensor product structure in which this condition would be fulfilled. This have been achieved in the following steps.

\begin{itemize}
\item We have introduced a notion of {\it maximally distant bases} (two bases interrelated through Fourier transform) and argued that the classical domain of the state space can not contain all vectors from two maximally distant bases.
\item We have pointed out that if the equilibrium state of an open system can be diagonal in either of two maximally distant bases depending on the initial conditions, this suffices to ensure that the decoherence-assisted classicality fails to emerge.
\item We have constructed a TPS within which the  above condition on the equilibrium state is fulfilled and thus the decoherence-assisted classicality (as well as thermalization) fails to emerge. In this TPS every eigenvector of the total Hamiltonian is of the product form, and if reduced to a state space of an open system, belongs to one of two maximally distant bases.
\end{itemize}

It is important to stress that the failure of the decoherence-assisted classicality and of the thermalization have been demonstrated for {\it generic} initial conditions, i.e. for all initial state vectors from large subspaces of the state space.

\section{Acknowledgements}

The author thanks David Steglet and Evgeniy Safonov for useful remarks.
The author is grateful to ERC (grant no. 279738 – NEDFOQ) for financial support. The partial support from grants
NSh-4172.2010.2, RFBR-11-02-00778 and RFBR-10-02-01398 is also acknowledged.
\vspace{1 mm}
%


\bibliography{D:/ITEP/QM/BibFiles/thermalization,D:/ITEP/QM/BibFiles/decoherence,D:/ITEP/QM/BibFiles/Lychkovskiy_partitioning_paper}

\begin{thebibliography}{29}%
\makeatletter
\providecommand \@ifxundefined [1]{%
 \@ifx{#1\undefined}
}%
\providecommand \@ifnum [1]{%
 \ifnum #1\expandafter \@firstoftwo
 \else \expandafter \@secondoftwo
 \fi
}%
\providecommand \@ifx [1]{%
 \ifx #1\expandafter \@firstoftwo
 \else \expandafter \@secondoftwo
 \fi
}%
\providecommand \natexlab [1]{#1}%
\providecommand \enquote  [1]{``#1''}%
\providecommand \bibnamefont  [1]{#1}%
\providecommand \bibfnamefont [1]{#1}%
\providecommand \citenamefont [1]{#1}%
\providecommand \href@noop [0]{\@secondoftwo}%
\providecommand \href [0]{\begingroup \@sanitize@url \@href}%
\providecommand \@href[1]{\@@startlink{#1}\@@href}%
\providecommand \@@href[1]{\endgroup#1\@@endlink}%
\providecommand \@sanitize@url [0]{\catcode `\\12\catcode `\$12\catcode
  `\&12\catcode `\#12\catcode `\^12\catcode `\_12\catcode `\%12\relax}%
\providecommand \@@startlink[1]{}%
\providecommand \@@endlink[0]{}%
\providecommand \url  [0]{\begingroup\@sanitize@url \@url }%
\providecommand \@url [1]{\endgroup\@href {#1}{\urlprefix }}%
\providecommand \urlprefix  [0]{URL }%
\providecommand \Eprint [0]{\href }%
\providecommand \doibase [0]{http://dx.doi.org/}%
\providecommand \selectlanguage [0]{\@gobble}%
\providecommand \bibinfo  [0]{\@secondoftwo}%
\providecommand \bibfield  [0]{\@secondoftwo}%
\providecommand \translation [1]{[#1]}%
\providecommand \BibitemOpen [0]{}%
\providecommand \bibitemStop [0]{}%
\providecommand \bibitemNoStop [0]{.\EOS\space}%
\providecommand \EOS [0]{\spacefactor3000\relax}%
\providecommand \BibitemShut  [1]{\csname bibitem#1\endcsname}%
\let\auto@bib@innerbib\@empty
\bibitem [{\citenamefont {Zeh}(1973)}]{zeh1973toward}%
  \BibitemOpen
  \bibfield  {author} {\bibinfo {author} {\bibfnamefont {H.}~\bibnamefont
  {Zeh}},\ }\href@noop {} {\bibfield  {journal} {\bibinfo  {journal}
  {Foundations of Physics}\ }\textbf {\bibinfo {volume} {3}},\ \bibinfo {pages}
  {109} (\bibinfo {year} {1973})}\BibitemShut {NoStop}%
\bibitem [{\citenamefont {Zeh}(1970)}]{zeh1970interpretation}%
  \BibitemOpen
  \bibfield  {author} {\bibinfo {author} {\bibfnamefont {H.}~\bibnamefont
  {Zeh}},\ }\href@noop {} {\bibfield  {journal} {\bibinfo  {journal}
  {Foundations of Physics}\ }\textbf {\bibinfo {volume} {1}},\ \bibinfo {pages}
  {69} (\bibinfo {year} {1970})}\BibitemShut {NoStop}%
\bibitem [{\citenamefont {Zurek}(1981)}]{zurek1981pointer}%
  \BibitemOpen
  \bibfield  {author} {\bibinfo {author} {\bibfnamefont {W.}~\bibnamefont
  {Zurek}},\ }\href@noop {} {\bibfield  {journal} {\bibinfo  {journal}
  {Physical Review D}\ }\textbf {\bibinfo {volume} {24}},\ \bibinfo {pages}
  {1516} (\bibinfo {year} {1981})}\BibitemShut {NoStop}%
\bibitem [{\citenamefont {Zurek}(1982)}]{zurek1982environment}%
  \BibitemOpen
  \bibfield  {author} {\bibinfo {author} {\bibfnamefont {W.}~\bibnamefont
  {Zurek}},\ }\href@noop {} {\bibfield  {journal} {\bibinfo  {journal}
  {Physical Review D}\ }\textbf {\bibinfo {volume} {26}},\ \bibinfo {pages}
  {1862} (\bibinfo {year} {1982})}\BibitemShut {NoStop}%
\bibitem [{\citenamefont {Schlosshauer}(2008)}]{schlosshauer2008decoherence}%
  \BibitemOpen
  \bibfield  {author} {\bibinfo {author} {\bibfnamefont {M.}~\bibnamefont
  {Schlosshauer}},\ }\href@noop {} {\emph {\bibinfo {title} {Decoherence and
  the Quantum-To-Classical Transition}}}\ (\bibinfo  {publisher} {Springer},\
  \bibinfo {year} {2008})\BibitemShut {NoStop}%
\bibitem [{\citenamefont {Tasaki}(1998)}]{tasaki1998quantum}%
  \BibitemOpen
  \bibfield  {author} {\bibinfo {author} {\bibfnamefont {H.}~\bibnamefont
  {Tasaki}},\ }\href@noop {} {\bibfield  {journal} {\bibinfo  {journal}
  {Physical review letters}\ }\textbf {\bibinfo {volume} {80}},\ \bibinfo
  {pages} {1373} (\bibinfo {year} {1998})}\BibitemShut {NoStop}%
\bibitem [{\citenamefont {Popescu}\ \emph {et~al.}(2006)\citenamefont
  {Popescu}, \citenamefont {Short},\ and\ \citenamefont
  {Winter}}]{popescu2006entanglement}%
  \BibitemOpen
  \bibfield  {author} {\bibinfo {author} {\bibfnamefont {S.}~\bibnamefont
  {Popescu}}, \bibinfo {author} {\bibfnamefont {A.}~\bibnamefont {Short}}, \
  and\ \bibinfo {author} {\bibfnamefont {A.}~\bibnamefont {Winter}},\
  }\href@noop {} {\bibfield  {journal} {\bibinfo  {journal} {Nature Physics}\
  }\textbf {\bibinfo {volume} {2}},\ \bibinfo {pages} {754} (\bibinfo {year}
  {2006})}\BibitemShut {NoStop}%
\bibitem [{\citenamefont {Goldstein}\ \emph {et~al.}(2006)\citenamefont
  {Goldstein}, \citenamefont {Lebowitz}, \citenamefont {Tumulka},\ and\
  \citenamefont {Zangh{\'\i}}}]{goldstein2006canonical}%
  \BibitemOpen
  \bibfield  {author} {\bibinfo {author} {\bibfnamefont {S.}~\bibnamefont
  {Goldstein}}, \bibinfo {author} {\bibfnamefont {J.}~\bibnamefont {Lebowitz}},
  \bibinfo {author} {\bibfnamefont {R.}~\bibnamefont {Tumulka}}, \ and\
  \bibinfo {author} {\bibfnamefont {N.}~\bibnamefont {Zangh{\'\i}}},\
  }\href@noop {} {\bibfield  {journal} {\bibinfo  {journal} {Physical review
  letters}\ }\textbf {\bibinfo {volume} {96}},\ \bibinfo {pages} {50403}
  (\bibinfo {year} {2006})}\BibitemShut {NoStop}%
\bibitem [{\citenamefont {Reimann}(2008)}]{reimann2008foundation}%
  \BibitemOpen
  \bibfield  {author} {\bibinfo {author} {\bibfnamefont {P.}~\bibnamefont
  {Reimann}},\ }\href@noop {} {\bibfield  {journal} {\bibinfo  {journal}
  {Physical review letters}\ }\textbf {\bibinfo {volume} {101}},\ \bibinfo
  {pages} {190403} (\bibinfo {year} {2008})}\BibitemShut {NoStop}%
\bibitem [{\citenamefont {Linden}\ \emph {et~al.}(2009)\citenamefont {Linden},
  \citenamefont {Popescu}, \citenamefont {Short},\ and\ \citenamefont
  {Winter}}]{linden2009quantum}%
  \BibitemOpen
  \bibfield  {author} {\bibinfo {author} {\bibfnamefont {N.}~\bibnamefont
  {Linden}}, \bibinfo {author} {\bibfnamefont {S.}~\bibnamefont {Popescu}},
  \bibinfo {author} {\bibfnamefont {A.}~\bibnamefont {Short}}, \ and\ \bibinfo
  {author} {\bibfnamefont {A.}~\bibnamefont {Winter}},\ }\href@noop {}
  {\bibfield  {journal} {\bibinfo  {journal} {Physical Review E}\ }\textbf
  {\bibinfo {volume} {79}},\ \bibinfo {pages} {061103} (\bibinfo {year}
  {2009})}\BibitemShut {NoStop}%
\bibitem [{\citenamefont {Gogolin}(2010)}]{gogolin2010pure}%
  \BibitemOpen
  \bibfield  {author} {\bibinfo {author} {\bibfnamefont {C.}~\bibnamefont
  {Gogolin}},\ }\emph {\bibinfo {title} {Pure State Quantum Statistical
  Mechanics}},\ \href@noop {} {Ph.D. thesis},\ \bibinfo  {school} {The
  University of Wurzburg} (\bibinfo {year} {2010}),\ \bibinfo {note}
  {arXiv:1003.5058}\BibitemShut {NoStop}%
\bibitem [{\citenamefont {Zanardi}(2001)}]{zanardi2001virtual}%
  \BibitemOpen
  \bibfield  {author} {\bibinfo {author} {\bibfnamefont {P.}~\bibnamefont
  {Zanardi}},\ }\href@noop {} {\bibfield  {journal} {\bibinfo  {journal}
  {Physical Review Letters}\ }\textbf {\bibinfo {volume} {87}},\ \bibinfo
  {pages} {077901} (\bibinfo {year} {2001})}\BibitemShut {NoStop}%
\bibitem [{\citenamefont {Gogolin}\ \emph {et~al.}(2011)\citenamefont
  {Gogolin}, \citenamefont {M{\"u}ller},\ and\ \citenamefont
  {Eisert}}]{gogolin2011absence}%
  \BibitemOpen
  \bibfield  {author} {\bibinfo {author} {\bibfnamefont {C.}~\bibnamefont
  {Gogolin}}, \bibinfo {author} {\bibfnamefont {M.}~\bibnamefont {M{\"u}ller}},
  \ and\ \bibinfo {author} {\bibfnamefont {J.}~\bibnamefont {Eisert}},\
  }\href@noop {} {\bibfield  {journal} {\bibinfo  {journal} {Physical Review
  Letters}\ }\textbf {\bibinfo {volume} {106}},\ \bibinfo {pages} {40401}
  (\bibinfo {year} {2011})}\BibitemShut {NoStop}%
\bibitem [{\citenamefont {Zurek}(1998)}]{zurek1998decoherence}%
  \BibitemOpen
  \bibfield  {author} {\bibinfo {author} {\bibfnamefont {W.}~\bibnamefont
  {Zurek}},\ }\href@noop {} {\bibfield  {journal} {\bibinfo  {journal}
  {Philosophical Transactions A}\ }\textbf {\bibinfo {volume} {356}},\ \bibinfo
  {pages} {1793} (\bibinfo {year} {1998})}\BibitemShut {NoStop}%
\bibitem [{\citenamefont {Omnes}(2002)}]{omnes2002decoherence}%
  \BibitemOpen
  \bibfield  {author} {\bibinfo {author} {\bibfnamefont {R.}~\bibnamefont
  {Omnes}},\ }\href@noop {} {\bibfield  {journal} {\bibinfo  {journal}
  {Physical Review A}\ }\textbf {\bibinfo {volume} {65}},\ \bibinfo {pages}
  {052119} (\bibinfo {year} {2002})}\BibitemShut {NoStop}%
\bibitem [{\citenamefont {Dugi{\'c}}\ and\ \citenamefont
  {Jekni{\'c}-Dugi{\'c}}(2012)}]{dugic2012parallel}%
  \BibitemOpen
  \bibfield  {author} {\bibinfo {author} {\bibfnamefont {M.}~\bibnamefont
  {Dugi{\'c}}}\ and\ \bibinfo {author} {\bibfnamefont {J.}~\bibnamefont
  {Jekni{\'c}-Dugi{\'c}}},\ }\href@noop {} {\bibfield  {journal} {\bibinfo
  {journal} {Pramana J Phys}\ }\textbf {\bibinfo {volume} {79}},\ \bibinfo
  {pages} {199} (\bibinfo {year} {2012})}\BibitemShut {NoStop}%
\bibitem [{\citenamefont {Deutsch}(1991)}]{deutsch1991quantum}%
  \BibitemOpen
  \bibfield  {author} {\bibinfo {author} {\bibfnamefont {J.}~\bibnamefont
  {Deutsch}},\ }\href@noop {} {\bibfield  {journal} {\bibinfo  {journal}
  {Physical Review A}\ }\textbf {\bibinfo {volume} {43}},\ \bibinfo {pages}
  {2046} (\bibinfo {year} {1991})}\BibitemShut {NoStop}%
\bibitem [{\citenamefont {Srednicki}(1994)}]{srednicki1994chaos}%
  \BibitemOpen
  \bibfield  {author} {\bibinfo {author} {\bibfnamefont {M.}~\bibnamefont
  {Srednicki}},\ }\href@noop {} {\bibfield  {journal} {\bibinfo  {journal}
  {Physical Review E}\ }\textbf {\bibinfo {volume} {50}},\ \bibinfo {pages}
  {888} (\bibinfo {year} {1994})}\BibitemShut {NoStop}%
\bibitem [{\citenamefont {Rigol}\ \emph {et~al.}(2008)\citenamefont {Rigol},
  \citenamefont {Dunjko},\ and\ \citenamefont
  {Olshanii}}]{rigol2008thermalization}%
  \BibitemOpen
  \bibfield  {author} {\bibinfo {author} {\bibfnamefont {M.}~\bibnamefont
  {Rigol}}, \bibinfo {author} {\bibfnamefont {V.}~\bibnamefont {Dunjko}}, \
  and\ \bibinfo {author} {\bibfnamefont {M.}~\bibnamefont {Olshanii}},\
  }\href@noop {} {\bibfield  {journal} {\bibinfo  {journal} {Nature}\ }\textbf
  {\bibinfo {volume} {452}},\ \bibinfo {pages} {854} (\bibinfo {year}
  {2008})}\BibitemShut {NoStop}%
\bibitem [{\citenamefont {Lieb}\ \emph {et~al.}(1961)\citenamefont {Lieb},
  \citenamefont {Schultz},\ and\ \citenamefont {Mattis}}]{lieb1961two}%
  \BibitemOpen
  \bibfield  {author} {\bibinfo {author} {\bibfnamefont {E.}~\bibnamefont
  {Lieb}}, \bibinfo {author} {\bibfnamefont {T.}~\bibnamefont {Schultz}}, \
  and\ \bibinfo {author} {\bibfnamefont {D.}~\bibnamefont {Mattis}},\
  }\href@noop {} {\bibfield  {journal} {\bibinfo  {journal} {Annals of
  Physics}\ }\textbf {\bibinfo {volume} {16}},\ \bibinfo {pages} {407}
  (\bibinfo {year} {1961})}\BibitemShut {NoStop}%
\bibitem [{\citenamefont {Abraham}\ \emph {et~al.}(1970)\citenamefont
  {Abraham}, \citenamefont {Barouch}, \citenamefont {Gallavotti},\ and\
  \citenamefont {Martin-L{\"o}f}}]{abraham1970thermalization}%
  \BibitemOpen
  \bibfield  {author} {\bibinfo {author} {\bibfnamefont {D.}~\bibnamefont
  {Abraham}}, \bibinfo {author} {\bibfnamefont {E.}~\bibnamefont {Barouch}},
  \bibinfo {author} {\bibfnamefont {G.}~\bibnamefont {Gallavotti}}, \ and\
  \bibinfo {author} {\bibfnamefont {A.}~\bibnamefont {Martin-L{\"o}f}},\
  }\href@noop {} {\bibfield  {journal} {\bibinfo  {journal} {Physical Review
  Letters}\ }\textbf {\bibinfo {volume} {25}},\ \bibinfo {pages} {1449}
  (\bibinfo {year} {1970})}\BibitemShut {NoStop}%
\bibitem [{\citenamefont {Lychkovskiy}(2011)}]{lychkovskiy2011entanglement}%
  \BibitemOpen
  \bibfield  {author} {\bibinfo {author} {\bibfnamefont {O.}~\bibnamefont
  {Lychkovskiy}},\ }\href@noop {} {\bibfield  {journal} {\bibinfo  {journal}
  {Journal of Physics: Conference Series}\ }\textbf {\bibinfo {volume} {306}},\
  \bibinfo {pages} {012028} (\bibinfo {year} {2011})}\BibitemShut {NoStop}%
\bibitem [{\citenamefont {Fel'dman}\ and\ \citenamefont
  {Zenchuk}(2012)}]{fel2012quantum}%
  \BibitemOpen
  \bibfield  {author} {\bibinfo {author} {\bibfnamefont {E.}~\bibnamefont
  {Fel'dman}}\ and\ \bibinfo {author} {\bibfnamefont {A.}~\bibnamefont
  {Zenchuk}},\ }\href@noop {} {\bibfield  {journal} {\bibinfo  {journal}
  {Physical Review A}\ }\textbf {\bibinfo {volume} {86}},\ \bibinfo {pages}
  {012303} (\bibinfo {year} {2012})}\BibitemShut {NoStop}%
\bibitem [{\citenamefont {Cucchietti}\ \emph {et~al.}(2005)\citenamefont
  {Cucchietti}, \citenamefont {Paz},\ and\ \citenamefont
  {Zurek}}]{cucchietti2005decoherence}%
  \BibitemOpen
  \bibfield  {author} {\bibinfo {author} {\bibfnamefont {F.}~\bibnamefont
  {Cucchietti}}, \bibinfo {author} {\bibfnamefont {J.}~\bibnamefont {Paz}}, \
  and\ \bibinfo {author} {\bibfnamefont {W.}~\bibnamefont {Zurek}},\
  }\href@noop {} {\bibfield  {journal} {\bibinfo  {journal} {Physical Review
  A}\ }\textbf {\bibinfo {volume} {72}},\ \bibinfo {pages} {052113} (\bibinfo
  {year} {2005})}\BibitemShut {NoStop}%
\bibitem [{\citenamefont {Lychkovskiy}(2010)}]{lychkovskiy2010necessary}%
  \BibitemOpen
  \bibfield  {author} {\bibinfo {author} {\bibfnamefont {O.}~\bibnamefont
  {Lychkovskiy}},\ }\href@noop {} {\bibfield  {journal} {\bibinfo  {journal}
  {Physical Review E}\ }\textbf {\bibinfo {volume} {82}},\ \bibinfo {pages}
  {011123} (\bibinfo {year} {2010})}\BibitemShut {NoStop}%
\bibitem [{\citenamefont {Zanardi}\ \emph {et~al.}(2004)\citenamefont
  {Zanardi}, \citenamefont {Lidar},\ and\ \citenamefont
  {Lloyd}}]{zanardi2004quantum}%
  \BibitemOpen
  \bibfield  {author} {\bibinfo {author} {\bibfnamefont {P.}~\bibnamefont
  {Zanardi}}, \bibinfo {author} {\bibfnamefont {D.}~\bibnamefont {Lidar}}, \
  and\ \bibinfo {author} {\bibfnamefont {S.}~\bibnamefont {Lloyd}},\
  }\href@noop {} {\bibfield  {journal} {\bibinfo  {journal} {Physical review
  letters}\ }\textbf {\bibinfo {volume} {92}},\ \bibinfo {pages} {60402}
  (\bibinfo {year} {2004})}\BibitemShut {NoStop}%
\bibitem [{\citenamefont {Haag}(1996)}]{haag1996local}%
  \BibitemOpen
  \bibfield  {author} {\bibinfo {author} {\bibfnamefont {R.}~\bibnamefont
  {Haag}},\ }\href@noop {} {\emph {\bibinfo {title} {Local Quantum Physics:
  Fields, Particles, Algebras}}},\ \bibinfo {edition} {2nd}\ ed.\ (\bibinfo
  {publisher} {Springer},\ \bibinfo {year} {1996})\BibitemShut {NoStop}%
\bibitem [{\citenamefont {Wallace}\ and\ \citenamefont
  {Timpson}(2010)}]{wallace2010quantum}%
  \BibitemOpen
  \bibfield  {author} {\bibinfo {author} {\bibfnamefont {D.}~\bibnamefont
  {Wallace}}\ and\ \bibinfo {author} {\bibfnamefont {C.}~\bibnamefont
  {Timpson}},\ }\href@noop {} {\bibfield  {journal} {\bibinfo  {journal} {The
  British Journal for the Philosophy of Science}\ }\textbf {\bibinfo {volume}
  {61}},\ \bibinfo {pages} {697} (\bibinfo {year} {2010})}\BibitemShut
  {NoStop}%
\bibitem [{\citenamefont {Joos}\ and\ \citenamefont
  {Zeh}(1985)}]{joos1985emergence}%
  \BibitemOpen
  \bibfield  {author} {\bibinfo {author} {\bibfnamefont {E.}~\bibnamefont
  {Joos}}\ and\ \bibinfo {author} {\bibfnamefont {H.}~\bibnamefont {Zeh}},\
  }\href@noop {} {\bibfield  {journal} {\bibinfo  {journal} {Z. Phys. B:
  Condensed Matter}\ }\textbf {\bibinfo {volume} {59}},\ \bibinfo {pages} {223}
  (\bibinfo {year} {1985})}\BibitemShut {NoStop}%
\end{thebibliography}%

\end{document}